 \documentclass[iop,apj]{emulateapj-rtx4}
\usepackage{apjfonts}
\usepackage{color}

\usepackage{amsmath,amssymb}
\usepackage{lscape}

\shorttitle{Compact starbursts in high-redshift SMGs}
\shortauthors{Ikarashi et al.}

\begin{document}


\title{Compact starbursts in $z$$\sim$3-6 submillimeter galaxies revealed by ALMA}


\author{Soh~Ikarashi\altaffilmark{1,2,3}, R.\,J.~Ivison\altaffilmark{1,4}, Karina\,I.~Caputi\altaffilmark{3}, Itziar~Aretxaga\altaffilmark{5}, James\,S.~Dunlop\altaffilmark{4}, Bunyo~Hatsukade\altaffilmark{6}, David\,H.~Hughes\altaffilmark{5}, Daisuke~Iono\altaffilmark{6,7}, Takuma~Izumi\altaffilmark{2}, Ryohei~Kawabe\altaffilmark{2,6,7}, Kotaro~Kohno\altaffilmark{2,8}, Claudia\,D.~P.~Lagos\altaffilmark{1}, Kentaro~Motohara\altaffilmark{2}, Kouichiro~Nakanishi\altaffilmark{6,7,9}, Kouji~Ohta\altaffilmark{10}, Yoichi~Tamura\altaffilmark{2}, Hideki~Umehata\altaffilmark{2}, Grant\,W.~Wilson\altaffilmark{11}, Kiyoto~Yabe\altaffilmark{6}, Min\,S.~Yun\altaffilmark{11} \\
{\scriptsize $^1${European Southern Observatory, Karl Schwarzschild Str.\ 2, D-85748 Garching, Germany}  \\
$^2${Institute of Astronomy, University of Tokyo, 2-21-1 Osawa, Mitaka, Tokyo 181-0015, Japan}  \\
$^3${Kapteyn Astronomical Institute, University of Groningen, P.O.\ Box 800, 9700 AV Groningen, The Netherlands; ${\rm sikarash@astro.rug.nl}$} \\
$^4${Institute for Astronomy, University of Edinburgh, Royal Observatory, Blackford Hill, Edinburgh EH9 3HJ, UK} \\
$^5${Instituto Nacional de Astrof\'{\i}sica, \'Optica y Electr\'onica (INAOE), Aptdo.\ Postal 51 y 216, 72000 Puebla, Mexico} \\
$^6${National Astronomical Observatory of Japan, Mitaka, Tokyo 181-8588, Japan} \\
$^7${SOKENDAI (The Graduate University for Advanced Studies), Shonan Village, Hayama, Kanagawa 240-0193, Japan}  \\
$^8${Research Center for the Early Universe, School of Science, University of Tokyo, 7-3-1 Hongo, Bunkyo, Tokyo 113-0033, Japan} \\
$^9${Joint ALMA Observatory, Alonso de Cordova 3107, Vitacura, Santiago 763 0355, Chile} \\
$^{10}${Department of Astronomy, Kyoto University, Kitashirakawa-Oiwake-Cho, Sakyo-ku, Kyoto 606-8502, Japan} \\
$^{11}${Department of Astronomy, University of Massachusetts, Amherst, MA 01003, USA}} }

\begin{abstract}
We report the source size distribution, as measured by ALMA millimetric continuum imaging, of a sample of 13 AzTEC-selected submillimeter galaxies (SMGs) at $z_{\rm phot}\sim3$--6.  Their infrared luminosities and star-formation rates (SFR) are $L_{\rm IR}\sim\,$2--$6\times 10^{12}$\,L$_{\odot}$ and $\sim200$--600\,M$_{\odot}$ yr$^{-1}$, respectively.  The sizes of these SMGs range from 0$''$.10 to 0$''$.38, with a median of 0$''$.20$^{+0''.03}_{-0''.05}$ (FWHM), corresponding to a median circularized effective radius ($R_{\rm c, e}$) of 0.67$^{+0.13}_{-0.14}$\,kpc, comparable to the typical size of the stellar component measured in compact quiescent galaxies at $z\sim 2$ (cQGs) --- $R_{\rm e}\sim 1$\,kpc. The median surface SFR density of our SMGs is 100$^{+42}_{-26}$\,M$_{\odot}$\,yr$^{-1}$\,kpc$^{-2}$, comparable to that seen in local merger-driven (U)LIRGs rather than in extended disk galaxies at low and high redshifts. The discovery of compact starbursts in $z\gtrsim3$ SMGs strongly supports a massive galaxy formation scenario wherein $z\sim3$--6 SMGs evolve into the compact stellar components of $z\sim2$ cQGs. These cQGs are then thought to evolve into the most massive ellipticals in the local Universe, mostly via dry mergers.  Our results thus suggest that $z\gtrsim3$ SMGs are the likely progenitors of massive local ellipticals, via cQGs, meaning that we can now trace the evolutionary path of the most massive galaxies over a period encompassing $\sim 90$\% of the age of the Universe.
\end{abstract}


\section{Introduction}

The most massive galaxies in the local Universe are thought to have evolved to their current state via a series of dry mergers of relatively gas-poor galaxies over the last 10\,Gyr \citep[e.g.][]{new12, ose12,car13,kro13}.  Their ancestors -- the so-called `compact quiescent galaxies’ (cQGs) -- are found at $z\sim 2$ in sensitive, near-infrared (NIR) imaging surveys \citep[e.g.][]{dad05, dok08, ono10, new12, kro13}.  These cQGs have $\sim 2$--$5\times$ smaller effective radii ($R_{\rm e}\sim 1$\,kpc) and are $\gtrsim 10\times$ denser than their local descendants \citep[e.g.][]{dok08, ono10, new12} and the process by which they form remains a mystery.  Recent attempts to probe their star-forming phase using conventional NIR observations resulted in the discovery of a relatively unobscured starburst, seen around $z\sim 2.5$--3 \citep{nel14, bar14}.  However, detailed simulations and population-synthesis modeling suggest that major mergers at $z\sim 3$--6 likely play a major role in the formation of the compact stellar component, via dust-obscured compact starbursts \citep[e.g.][]{wuy10, tof14}.  We must thus penetrate deep within these dusty environments to reveal this vigorous starburst phase.

Submillimeter galaxies (SMGs) \citep[e.g.][]{sib97, ivi98, hug98} have long been thought to be plausible progenitors of massive passive galaxies around $z\sim1.5$--2 based on their volume densities \citep{bla04, cha05}.  Early source size measurements for $z\sim1$--3 SMGs -- using radio continuum and CO emission-line data -- reported a median source size of $\sim 0''$.5 (full width at half maximum; FWHM) corresponding to a radius of $\sim2$--3\,kpc \citep[e.g.][]{tac06, big08}.  These early studies resulted in the common notion that high-redshift SMGs have larger star-forming regions than luminous, dusty galaxies in the local Universe, indicating that the size of their star-forming region is inconsistent with the compact structure of cQGs \citep[although a few SMGs at $z\sim2$ with compact cores were reported --][]{tac08}.  The size of starburst regions in SMGs at $z\gtrsim 3$ has remained largely unexplored, partly because it is difficult to identify SMGs at $z\gtrsim 3$, partly because the cosmological dimming then makes it difficult to measure their source sizes in the radio regime.  There have been a few source size measurements via (sub)millimeter continuum imaging for $z\gtrsim 3$ SMGs; SMA observations of AzTEC1 at $z\sim 4$ and PdBI observations of HFLS3 at $z=6.3$ revealed radii of $\sim 1.3$\,kpc \citep{you08, rie13}.  However these two are among the most brightest submillimeter galaxies known ($L_{\rm IR}>10^{13}$\,L$_{\odot}$) and we need to image more typical SMGs with $L_{\rm IR}\sim 10^{12}$\,L$_{\odot}$.

Sensitivity limitations of existing arrays meant we needed to wait for ALMA in order to measure the far-infrared (FIR) sizes of $z\gtrsim3$ SMGs for a significant sample of targets.  Here, we exploit high-resolution continuum imaging with ALMA to peer within a carefully selected sample of the most distant SMGs -- the most vigorous, dust-obscured, starburst galaxies in the early Universe.  We demonstrate that they have the compact starburst nuclei necessary to produce the small structures that typify cQGs.  We adopt throughout a cosmology with $H_{\rm 0}= 70$\,km\,s$^{-1}$\,Mpc$^{-1}$, $\Omega_{\rm M}= 0.3$ and $\Omega_{\rm \Lambda}= 0.7$.

\section{AzTEC-selected $z\gtrsim 3$ SMGs} 

Our ALMA program (2012.1.00326.S,\,P.I.\,Ikarashi; Ikarashi et al.\,in prep) was designed to study the most distant dusty starbursts, for which redshift estimates were obtained based on (sub)millimeter/radio \citep{car99} and red (sub)millimeter \citep[e.g.][]{rie13,hug02} colors.  
In this ALMA program,  we observed 30 AzTEC sources in the Subaru/XMM-Newton Deep field (SXDF), which includes the UKIDSS UDS field \citep{hat11,ika13}. The AzTEC 1100-$\mu$m map contains a total of 221 millimeter sources over a contiguous area of 950 arcmin$^2$. We selected our ALMA targets  based on their faintness in the $\it Herschel$ images \citep[$S_\nu (250 \, \rm  \mu m)< 18.3 \, \rm mJy\,beam^{-1}$; 3$\sigma$; ][]{oli12} and VLA 1.4-GHz map ($\lesssim$35 $\mu$Jy; 5$\sigma$) (Arumugam et al., in preparation). With the ALMA observations we detected 35 significant ($\geq$ 5$\sigma$) SMGs (hereafter ASXDF sources) associated with 30 AzTEC sources. 

Given the strong negative K-correction at $\lambda\sim$800--1300\,$\mu$m, the faintness in the {\it Herschel} and radio bands indicate that these 1100-$\mu$m-selected galaxies are expected to be at high redshifts, i.e., $z\gtrsim3$. Note that these galaxies constitute a complementary population to that studied in early sub-millimetre galaxy studies, which were biased towards radio-bright sources, and found to lie at lower redshifts $z\approx 1$--3 \citep[e.g.][]{cha05}.

Among the 35 ALMA sources, a total of 17 have detections with $S/N \geq10$ in the ALMA 1100-$\mu$m continuum map. Such a high $S/N$ threshold ensures that we can study their sizes with good accuracy with the ALMA continuum data (See details in Section\,3.1). As the focus of this paper is on $z\gtrsim3$ SMGs, we analyse here only the 13 (out of 17) sources that have photometric redshifts $z_{\rm phot}\geq2.8$, or are faint in IRAC ($F_{\rm 4.5\,\mu m}\geq$ 22.75 m$_{\rm AB}$) and detected in at most four optical/near-mid-IR broad bands, indicating a likely high redshift, as we explain below. 

At these shorter wavelengths (optical through mid-IR), we have performed the spectral energy distribution (SED) analysis of our sources based on 12 bands, namely  $B$, $V$, $Rc$, $i$, $z'$, $J$, $H$, $Ks$-bands and IRAC 3.6, 4.5, 5.8 and 8.0\,$\mu$m (Ikarashi et al.\,in prep), using the same method described in \citet{cap12}. For three out of our 13 sources, we obtained redshift estimates, $z_{\rm phot}$, and derived parameters (Table~\ref{tbl-1}). The remaining ten sources are only detected in four or less broad bands, so no robust $z_{\rm phot}$ can be obtained from the SED fitting.  

\begin{figure}
\begin{center}
\includegraphics[angle=0,scale=.4]{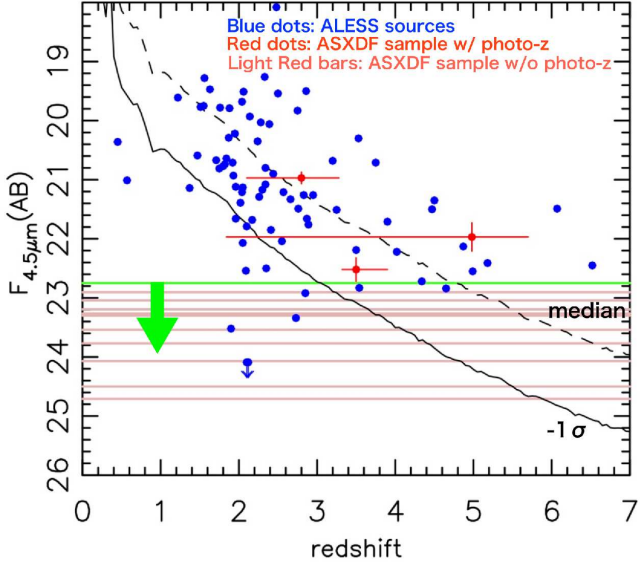}
\vspace{2.0mm}
\caption{Observed 4.5-$\mu$m flux of submillimeter galaxies as a function of redshift. Blue dots mark ALMA-identified LABOCA (ALESS) sources \citep{sim14}. Red points mark ASXDF sources with photometric redshifts in our sample for source size measurements. 
Black curves show the redshift-4.5-$\mu$m relation expected from the absolute $H$-band flux distribution of ALESS sources and the optical/NIR SED of average ALESS sources \citep{sim14}; Dashed line is for SMGs with the median absolute $H$-band flux; solid shows the absolute $H$-band flux distribution minus $1\sigma$.   
Light red bars mark the 4.5-$\mu$m flux of the ASXDF sources without photometric redshifts.  
Solid green horizontal bar marks $F_{4.5\,\mu m}$ at 22.75 m$_{\rm AB}$, which is the threshold for selection of $z\gtrsim3$ SMGs in this paper. 
In this paper, we adopt the cross points between the solid black curve and the horizontal light red lines as the expected 1$\sigma$ lower limits of redshift for each source (these values are listed in Table~\ref{tbl-1}). \label{fig:z45mic}}
\end{center}
\end{figure}

Figure~\ref{fig:z45mic} shows a 4.5-$\mu$m--redshift plot for ALMA sources and those in the ALMA-identified SMG sample (ALESS) reported by \citet{sim14}. The dashed line in this plot indicates the median  4.5-$\mu$m--redshift relation, and the solid line corresponds to this same relation minus the $1\sigma$ scatter, which we have derived using the average SED of the ALESS sources \citep[see Figure~8 in][]{sim14}.  We expect more than 85\% of the ASXDF sources with  $F_{\rm 4.5\,\mu m} \geq22.75$ m$_{\rm AB}$ to be located at $z\gtrsim3$. About 15\% of SMGs are expected to have  $F_{\rm 4.5\,\mu m}$ fainter than the solid black curve in Figure~\ref{fig:z45mic} at each redshift, and 15\% of SMGs at $z=3$ are expected to have $F_{\rm 4.5\,\mu m} \geq22.75$ m$_{\rm AB}$. So, by selecting only those galaxies with no redshift estimate in our sample, with $F_{\rm 4.5\,\mu m} \geq22.75$, we obtain a conservative list of sources likely to lie at $z\gtrsim 3$. In Table~\ref{tbl-1},  we list the expected minimum redshifts for our sources based on the solid line in Figure~\ref{fig:z45mic}. 

\begin{figure}
\begin{center}
\includegraphics[angle=0,scale=.35]{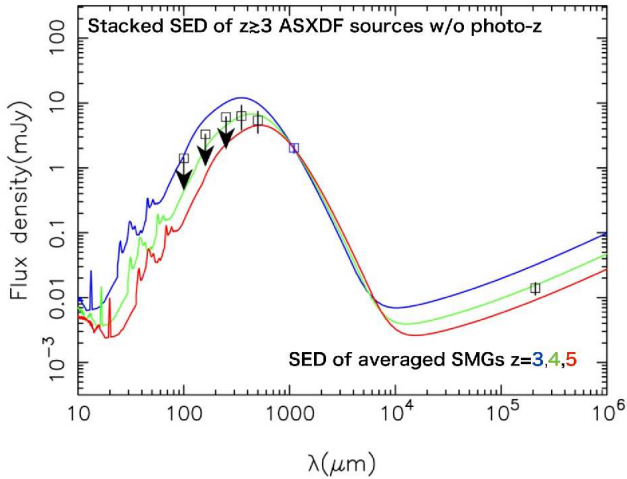}
\vspace{2.0mm}
\caption{Stacked submillimeter/radio SED of the ASXDF sources without photo-$z$. Error are estimated by Bootstrapping analysis.  
Colored SED is that of average SMGs \citep{swi14} for $z=$3, 4 and 5, as best fit to the ALMA flux. For PACS 100 and 160 $\mu$m and SPIRE 250 $\mu$m data, we plot 3$\sigma$ upper limits. \label{fig:sed}}
\end{center}
\end{figure}

The stacked submillimeter--radio SED of these optical/NIR dropout SMGs also help us to understand, in an independent manner, whether our galaxies are actually located at $z\gtrsim3$. 
Figure~\ref{fig:sed} shows the stacked fluxes at 100, 160, 250, 350, 500\,$\mu$m (PACS and SPIRE), 1100\,$\mu$m (ALMA) and 21\,cm (VLA) with the SED of the averaged SMGs ($T_{\rm d}=32$\,K) at $z=$3, 4 and 5.  All of the stacked fluxes and errors are based on bootstrapping analysis. We see that the stacked submillimeter/radio fluxes are best fitted at $z\sim4$.  Given a stacked VLA 1.4-GHz flux density of $15.2\pm 2.4$\,$\mu$Jy and the observed ALMA 1100-$\mu$m flux density, we expect a photometric redshift, $z=4.0^{+0.4}_{-0.4}$, for our ASXDF sources, based on their radio/(sub)millimeter color \citep[e.g.][]{car99}.  Note that, for this exercise, we have considered the radio-FIR SED template of the averaged SMGs \citep{swi14} derived from the ALMA-identified SMGs at $z\sim2$ \citep{sim14}. Here we are assuming that this template is also valid at $z\gtrsim3$--4.  If $T_{\rm d}$ follows the trend that SMGs at higher redshifts have higher $T_{\rm d}$, then the redshifts should indeed be $z\gtrsim4$. At the moment, it is difficult to be confident of such a trend as the samples of known $z\gtrsim3$--4 SMGs are small.  Nevertheless, from all our arguments based on the multi-wavelength SED study of our galaxies from the optical through the radio, we can conclude that our 13 ASXDF/ALMA sources are safe candidates for $z\gtrsim3$ SMGs.

Our derived flux densities at 1100-$\mu$m range from 1.5 to 3.4\,mJy, corresponding to star formation rates (SFRs) of $\sim 200$--600\,M$_{\odot}$\,yr$^{-1}$ and $L_{\rm IR}\sim 2$--$6\times 10^{12}$\,L$_{\odot}$.  The median SFR and $L_{\rm IR}$ are 340$^{+12}_{-13}$\,M$_{\odot}$\,yr$^{-1}$ and  3.4$^{+0.1}_{-0.1}\times$10$^{12}$\,L$_{\odot}$, respectively. The properties of our sample are summarized in Table~\ref{tbl-1}. The SFRs and $L_{\rm IR}$ are estimated from the average SMG SEDs.  We considered uniform redshift probability density at $z=3-6$ for sources without a $z_{\rm phot}$ determination, and a 1$\sigma$ error for sources which do have a  $z_{\rm phot}$.

\section{Source-size measurements} 

\subsection{Data, method and results}

\begin{figure}
\begin{center}
\includegraphics[angle=0,scale=.50]{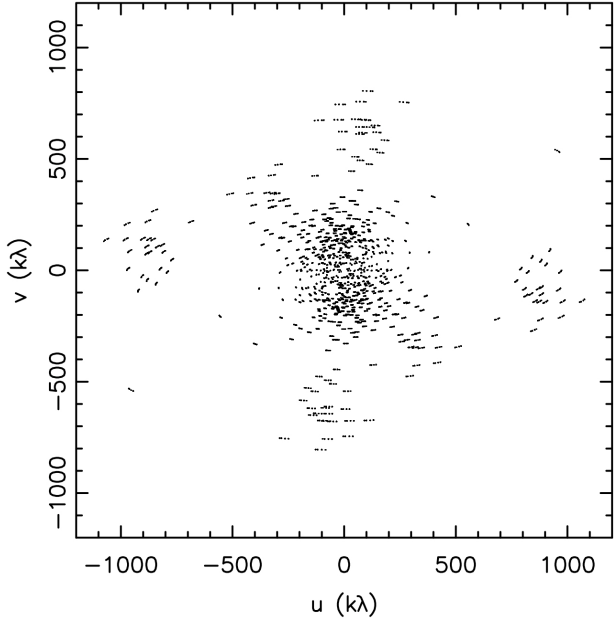}
\vspace{2.0mm}
\caption{The $uv$ coverage for ASXDF sources with long-baseline antennas for Schedule Block 1 in Table.~\ref{tbl-1}. We use visibilities at $uv$ distances of $\leq400$\,k$\lambda$ for source-size measurements where $u$--$v$ coverage is well sampled. 
We use visibilities at 400--1200\,k$\lambda$ only to check the consistency between the expected long-baseline visibilities and the measured size.  \label{fig:uvcov}}
\end{center}
\end{figure}

We measured the source sizes of our ASXDF sources using ALMA continuum data centered at 265\,GHz.  Our ALMA observations were obtained in three blocks, with only small differences in antenna configurations between blocks.  Seven of the 13 ASXDF sources were observed with 25 working 12-m antennas, mainly covering baselines up to 400\,k$\lambda$, corresponding to physical baseline lengths of 440\,m.  The remaining six sources were observed with three more 12-m antennas, deployed for tests on longer baselines, covering up to 1200\,k$\lambda$ or 1320\,m.  The extended-baseline data from 400 to 1200\,k$\lambda$ are used here only as supplementary data because of their limited $uv$ coverage (Figure~\ref{fig:uvcov}). On-source observation times were 3.6--4.5 minutes, sufficient to achieve r.m.s.\ noise levels of 70--88\,$\mu$Jy\,beam$^{-1}$. The synthesized beam size in our ALMA continuum images, using baselines up to 400\,k$\lambda$, is $\sim 0''.7$ (FWHM) -- too coarse to allow us to resolve any compact starburst nuclei in high-redshift SMGs. All of our sample shows the millimeter sizes of $\gtrsim2$ times smaller than the beam size by a CASA task, {\sc imfit}; about the 5 of the 13 are unresolved or point-like. 

In this paper we have measured source sizes using the visibility data directly -- on $uv$--amplitude plots (hereafter $uv$-amp plots) -- assuming a symmetrical Gaussian\footnote{If we measure the size of a disk-like source using a Gaussian fit, the actual size of the disk ($R_{\rm e, Disk}$) is empirically $\sim 1.1\times$ larger than the size measured ($R_{\rm e, Gauss}$) for measurements at $\leq400$\,k$\lambda$.} as was done in previous studies, to exploit the long-baseline ($\leq$400\,k$\lambda$) data for source size measurements (Figure~\ref{fig:uvamp}).  Source-size measurements using $uv$-amp plots have often been made in previous studies using, e.g., SMA and CARMA, in order to better constrain the size of largely unresolved sources in an image \citep[e.g.][]{ion06, you08, ivi10, ika11}.  This is equivalent to measuring the circularized effective radius, $R_{\rm c, e}$.

In this paper, we have been able to polish this method, owing to the high data quality from ALMA.  We have evaluated the accuracy of our source-size measurements using a Monte-Carlo simulation, for the purpose of correcting for any systematics and obtaining more reliable source sizes.  We generated 82000 mock sources with a symmetric Gaussian profile in noisy visibility data, for a range of source sizes and flux densities that cover the putative parameter range of our ASXDF sources.  We measured source sizes and created cleaned continuum images in the same manner as we had done for our real targets, in order to derive a relation between the input source size, the measured source size and the signal-to-noise ratio in a continuum image.  Figure~\ref{fig:simu} shows the derived relation between measured source size found by fitting in $uv$-amp plots and the `actual' size input for the simulation, each versus source size for continuum detections of 10 and 15$\sigma$.  This plot demonstrates that our source size measurement is accurate to within 1$\sigma$ and that actual source sizes are systematically a little bit larger than the measured source sizes.  We therefore adopt source sizes after making a correction based on this relation between measured and actual source sizes; the correction is done using the probability distribution of actual source size for the appropriate signal-to-noise ratio and measured size of each source.  
In this paper, we measure millimeter size of ASXDF sources with $\geq10\sigma$ continuum detections. 
This is because size measurement in the visibility data for 10$\sigma$ sources gets less sensitive at FWHM\,$<$0$''$.2, i.e., losing inearity (Figure~\ref{fig:simu}). 
Measurement of our sample is safe from this issue; there is just one ASXDF source with S/N$=$10, but its measured size is $\sim$0$''$.3, and we checked that our measurement for the second lowest S/N of 11.3, is sensitive down to 0$''$.1. 
 
The measured source sizes are listed in Table~\ref{tbl-1}.  The source sizes of our sample range from 0$''$.10 to 0$''$.38 with a median of 0$''$.20$^{+0''.03}_{-0''.05}$ (Figure~\ref{fig:size}).  We also check the dependency of the circularized size measured in $uv$-amp plots on ellipticity via simulations, for minor/major axis ratio of 0.5, 0.6, 0.7, 0.8, 0.9 and 1.0 (Figure~\ref{fig:simu}).  These simulation suggest that the measured circularized size does not depend strongly on the ellipticity; however, a weak dependency exists: at an axis ratio of 0.5, the circularized size can be over-estimated about 10 percent.

\begin{figure*}
\begin{center}
\includegraphics[angle=0,scale=0.68]{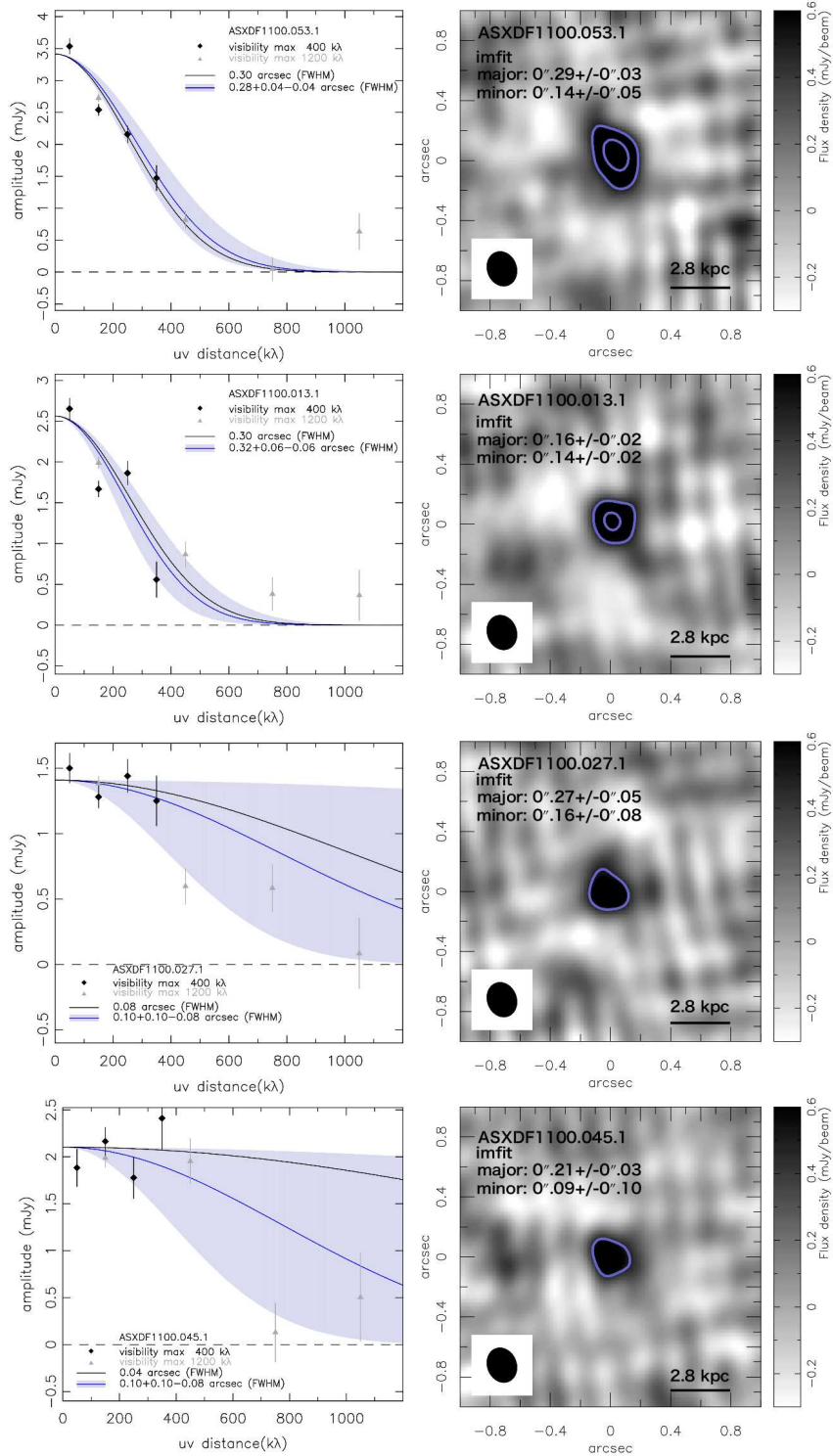}
\vspace{0.0mm}
\caption{Size measurements for six of the sources with long-baseline data (400--1200 k$\lambda$) in our sample and stacked visibility data. 
Stacked (all) includes all of ASXDF sources with long-baseline data (ASXDF1100.013.1, 27.1, 45.1, 45.2, 49.1 and 53.1). 
Stacked (faint) includes faint ASXDF sources with long-baseline data (ASXDF1100.027.1, 45.1, 45.2 and 49.1).
({\it Left:}) Black and grey points show the $uv$ visibilities up to 400 and 1200 k$\lambda$, respectively. 
The black line is a $uv$-amp model of the best-fitted Gaussian component. 
The blue line and shaded area are possible solutions for the corrected source size, with errors listed in Table~\ref{tbl-1}.  
The blue line and shaded area are plotted for the total amplitude of the best-fitted model. 
({\it Right:}) ALMA 1100-$\mu$m continuum images with synthesized beam sizes of $\sim$ 0$''$.2, generated by using 200--1200 k$\lambda$ data. 
The r.m.s.\ in images of ASXDF1100.053.1, 13.1, 27.1, 45.1, 45.2, 49.1 and {\it stacked faint} and {\it stacked all} are 126, 124, 124, 122, 122, 126, 88 and 67 $\mu$Jy\,beam$^{-1}$, respectively.  Contours are shown at +4$\sigma$ and +8$\sigma$.   
These $uv$-amp plots and high-angular-resolution images using $\leq$ 1200-k$\lambda$ data imply that these sources have a single, compact component, as shown by our source size analysis using $\leq 400$-k$\lambda$ data. 
\label{fig:uvamp}}
\end{center}
\end{figure*}

\addtocounter{figure}{-1}
\begin{figure*}
\begin{center}
\includegraphics[angle=0,scale=0.68]{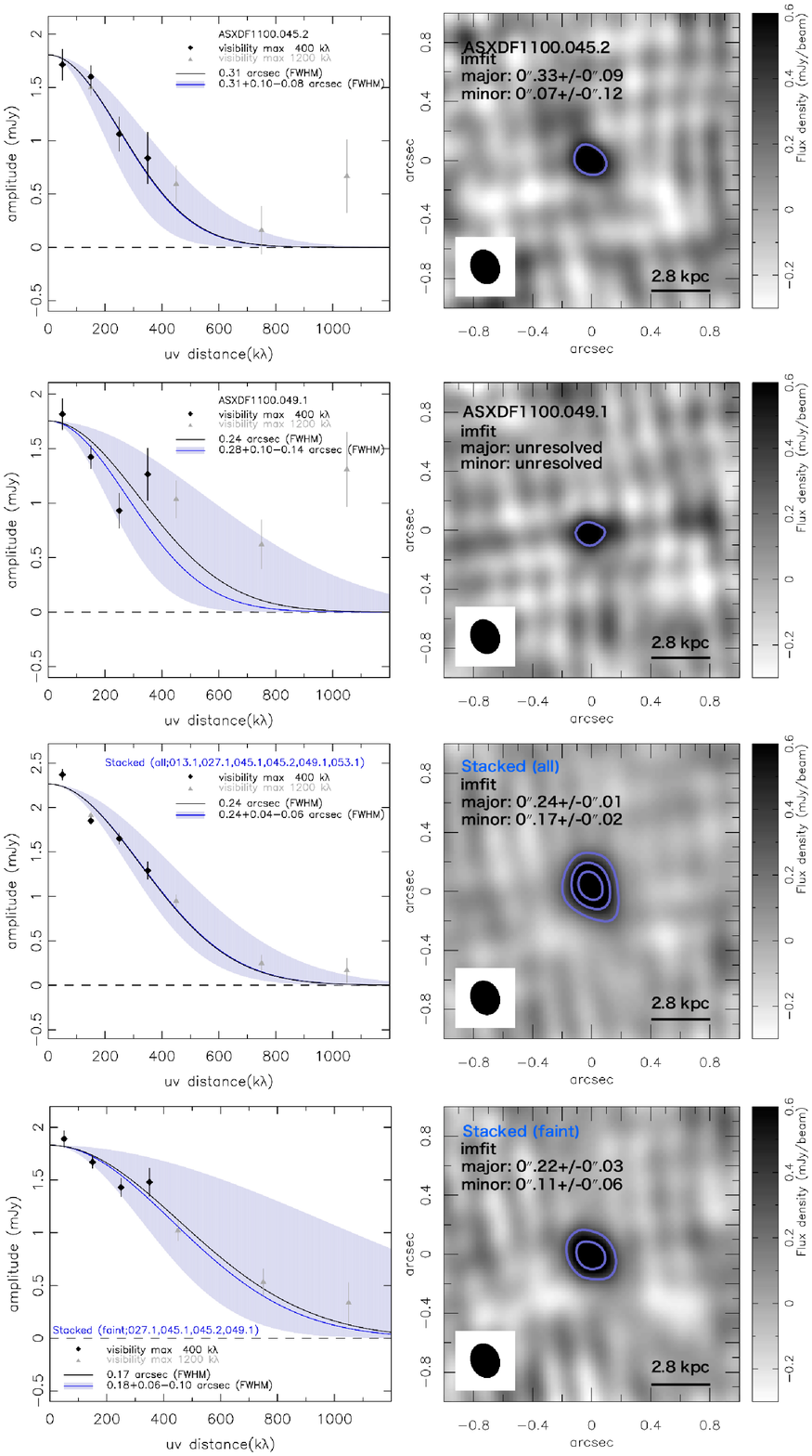}
\vspace{0.0mm}
\caption{Continued \label{fig:uvamp}}
\end{center}
\end{figure*}

\subsection{Ancillary long-baseline data}

As we noted above, six of 13 ASXDF sources in our sample were observed with additional three long-baseline antennas covering up to 1200\,k$\lambda$ (Table~\ref{tbl-1}) which enable us to make millimeter images of the sources with an angular resolution of 0$''$.2 (FWHM).  Using the long-baseline data, we check some concerns in our source size measurements --- assumptions made about source multiplicity (i.e.\ that there is none), the possibility of source ellipticity, and the possibility of faint, extended emission. In Figure~\ref{fig:uvamp}, we present high-resolution ALMA continuum images and $uv$-amp plots of ASXDF1100.013.1, 27.1, 45.1, 45.2, 49.1 and 53.1, which are covered by long-baseline data individually.  Moreover in order to check the properties of fainter sources with better signal-to-noise --- ASXDF1100.027.1, 045.1, 045.2 and 049.1 --- we stacked the visibility data of these four sources using the CASA code, {\sc stacker} \citep{lin15}.  We also stacked the six sources with long-baseline data to check their average properties.  Hereafter we refer to the former and latter stacked data as {\it stacked faint} and {\it stacked all}.

First, $uv$-amp plots of ASXDF1100.013.1, 27.1, 45.1, 45.2, 49.1 and 53.1 and {\it stacked faint} and {\it stacked all} demonstrate that estimating their size by $uv$-amp analysis using up to 400\,k$\lambda$ yields results consistent with their long-baseline visibility data up to 1200\,k$\lambda$ (Figure~\ref{fig:uvamp}).

Second, we created high-angular-resolution ALMA millimeter continuum images (hereafter high-res images) using 200--1200\,k$\lambda$ baselines (Figure~\ref{fig:uvamp}).  The resultant synthesized beam is 0$''$.23 $\times$ 0$''$.19 (PA $=$ 21$^{\circ}$).  The r.m.s.\ of the images of ASXDF1100.053.1, 13.1, 27.1, 45.1, 45.2, 49.1, {\it stacked faint} and {\it stacked all} are 126, 124, 124, 122, 122, 126, 88 and 67 $\mu$Jy beam$^{-1}$, respectively.  The sources are detected with $S/N_{\rm peak} =$11, 9, 7, 7, 6, 6, 10 and 15$\sigma$, respectively.  We measured their millimeter sizes and fluxes in the image using a CASA task, {\sc imfit}: 0$''$.16-0$''$.29 (major axis).  As suggested by our source size measurements via $uv$-amp fitting, each of our $z\gtrsim3$ SMGs has a compact star-forming region.

Next, we check whether or not the compact star-forming region dominates the huge SFRs of these SMGs.
 Usually we would simply compare the flux of a compact component on a high-res image with the total flux in order to estimate the flux fraction emitted by the compact component.   However, we remove the ALMA data at $uv$-distances $<200$\,k$\lambda$ in order to sharpen the synthesized beam, so we need to take into account any missing flux. 
In this paper, then, we compare fluxes measured using {\sc imfit} on the high-res ALMA images with fluxes expected at a $uv$-distance of $\geq200$\,k$\lambda$. 
We adopt the fluxes measured at a $uv$-distance of $200$\,k$\lambda$ (Figure~\ref{fig:uvamp}) as the flux expected for the measured size. 
Figure~\ref{fig:comp} shows the comparison between the flux measured at a $uv$-distance of $\geq200$\,k$\lambda$ and flux measured via {\sc imfit}. 
The comparison shows that the fluxes measured by {\sc imfit} are almost the same as the fluxes measured using the visibilities. 
The relation between fluxes from the image and from the visibility data can be fit by
\begin{equation}
F_{\rm image, 200\,k\lambda} = 1.07^{+0.08}_{-0.09} \times F_{\rm visibility, 200\,k\lambda}, 
\end{equation}
where $F_{\rm image, 200\,k\lambda}$ is the flux measured by {\sc imfit} and $F_{\rm visibility, 200\,k\lambda}$ is that measured from the visibilities.  This indicates that $\sim$\,100 percent of the rest-frame FIR emission in $z\gtrsim3$ ASXDF sources comes from the compact component shown in the high-res images.

Last, we check the ellipticity of sources. 
Sizes measured by {\sc imfit} are shown on the high-res ALMA images in Figure~\ref{fig:uvamp}, and tend to show ellipticity, i.e., minor/major axis ratio of a fitted asymmetric Gaussian $<$ 1; the minor/major axis ratio of ASXDF1100.053.1, 013.1, 027.1, 045.1 and 045.2 are 0.5$\pm$0.2, 0.9$\pm$0.2, 0.6$\pm$0.3, 0.4$\pm$0.5, and 0.2$\pm$0.4, respectively. 
Here we check the ellipticity shown in the individual sources with  $S/N_{\rm peak}\sim$10  (Figure~\ref{fig:uvamp}). 
In order to test whether the elliptical feature is real, we checked the empirical accuracy of {\sc imfit} using Monte Carlo simulations for major/minor axis ratios of 0.5, 0.6, 0.7, 0.8, 0.9 and 1.0 and circularized sizes of 0$''$.20, 0$''$.30 and 0$''$.40 (FWHM) and $S/N_{\rm peak}\sim10$ (Figure~\ref{fig:isimu}).  
The simulations show two features. 
First, {\sc imfit} tends to return major/minor axis ratio of $<1$ even for input axis ratio of 1. 
Second, {\sc imfit} also tends to report smaller sizes for larger input circularized sizes such as 0$''$.30 and 0$''$.40. 
The latter is partly because of missing flux in our high-res images and partly because their detection of  $S/N_{\rm peak} \sim10$ is too low to fit the extended emission. 
In light of these simulations, the sizes and ellipticities measured by {\sc imfit} on the stacked images are not inconsistent with symmetric Gaussian emission (minor/major axis ratio $=$ 1). 
Given the measured ellipticity for the {\it stacked all} and {\it stacked faint} are 0.7 and 0.5, respectively, and their $R_{\rm c, e}$ are 0$''$.24 and 0$''$.17, respectively, the simulations indicate that they can in fact be symmetric Gaussians. 
When we see the input-output ellipticity plot for $R_{\rm c, e}=$\,0$''$.20 (bottom in Figure~\ref{fig:isimu}), the measured ellipticity for the {\it stacked faint} looks off 1$\sigma$ error of the ellipticity of 1 but with in 1.4$\sigma$. 

Next, we investigate individual ASXDF sources with $\sim$ 10$\sigma$ detections in the high-res ALMA image, ASXDF1100.053.1 and 013.1. 
ASXDF1100.053.1 has a size estimated via its visibility data of 0$''$.28 and shows an ellipticity of 0.5 via {\sc imfit}. 
Because of our simulations we cannot exclude the possibility of symmetric Gaussian emission in ASXDF1100.053.1, but an ellipticity of $\lesssim0.7$ seems to be more plausible. 
According to our simulations for $R_{\rm c, e}=$\,0$''$.30 (bottom in Figure~\ref{fig:isimu}), the measured ellipticity of ASXDF1100.053.1 has a probability of only 1.3 per cent that ASXDF1100.053.1 has ellipticity of $\geq0.8$. Thus ASXDF1100.053.1 has ellipticity of  $\lesssim0.7$ plausibly. 
Moving on to ASXDF1100.013.1, there is not a mock source with the size of ASXDF1100.013.1, 0$''$.16$\times$0$''$14, in the simulation for an input circularized size of 0$''$.30. This implies that ASXDF1100.013.1 may have a starburst region more concentrated in the center than a Gaussian profile.  
These shapes that are unlikely to be symmetric Gaussians may be giving us a hint of complex  star-forming structure in the small emission area. 

\begin{figure}
\begin{center}
\includegraphics[angle=0,scale=.55]{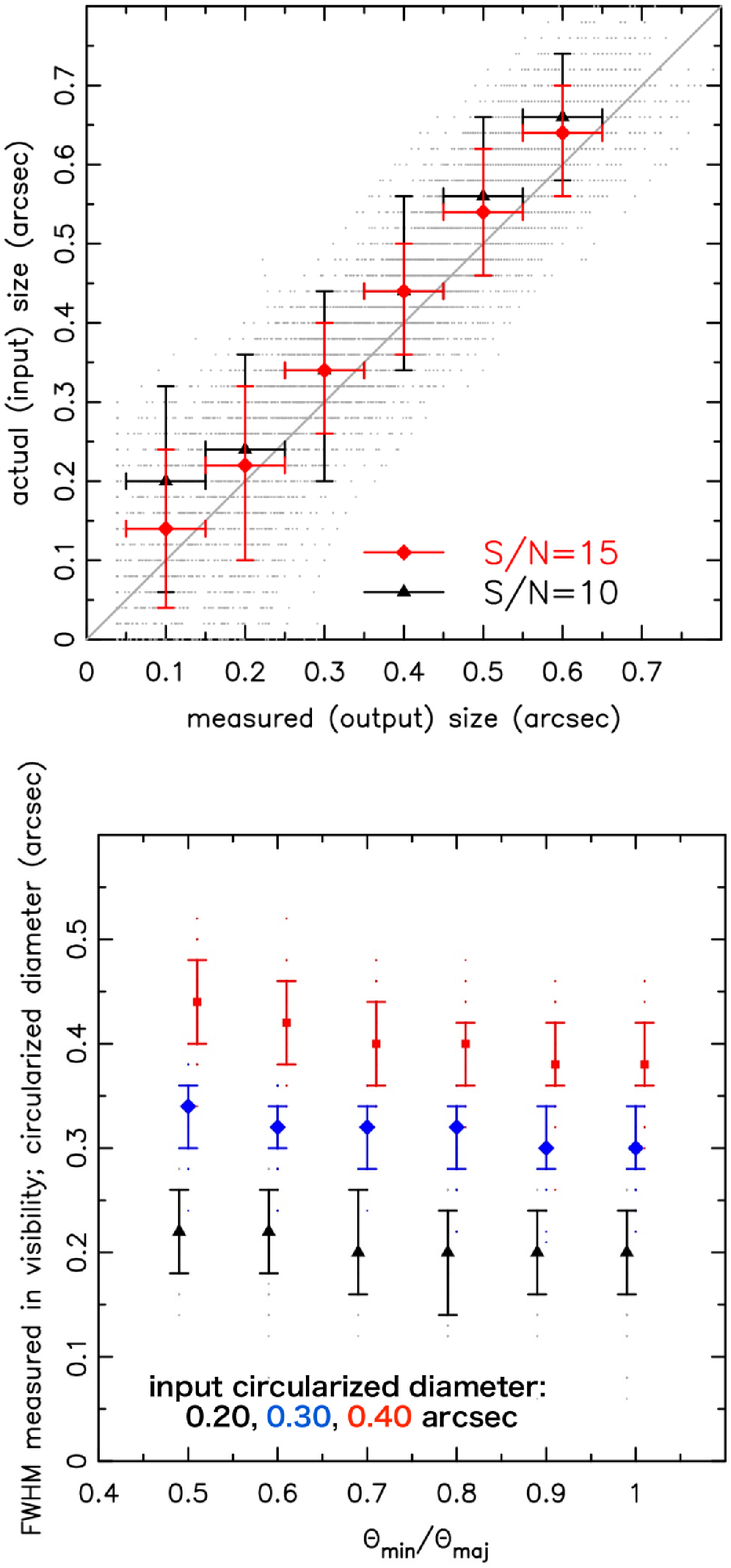}
\vspace{8.0mm}
\caption{(Top) Relationship between `raw' measured sizes from fitting in $uv$-amp plot and `actual' sizes derived by our Monte Carlo simulation in noise visibility data for ALMA sources with 10 and 15$\sigma$ ALMA continuum detections. Grey dots mark mock sources with 15$\sigma$ detections. 
This plot shows how the input size for mock sources compares with the size measured by fitting to the $uv$-amp plot. Error bars show 1$\sigma$ for the input source size distributions.  The plot indicates that measurements for low signal-to-noise sources get less effective at $\lesssim$0$''$.10, requiring larger corrections. (Bottom) Relationship between intrinsic minor/major axis ratio and measured circularized size by $uv$-amp plot based on another simulation. We plot adding offset of $-$0.01, 0 and $+$0.01 to $\theta_{\rm min}$/$\theta_{\rm maj}$ for visibility. \label{fig:simu}}
\end{center}
\end{figure}

\begin{figure}
\begin{center}
\includegraphics[angle=0,scale=.55]{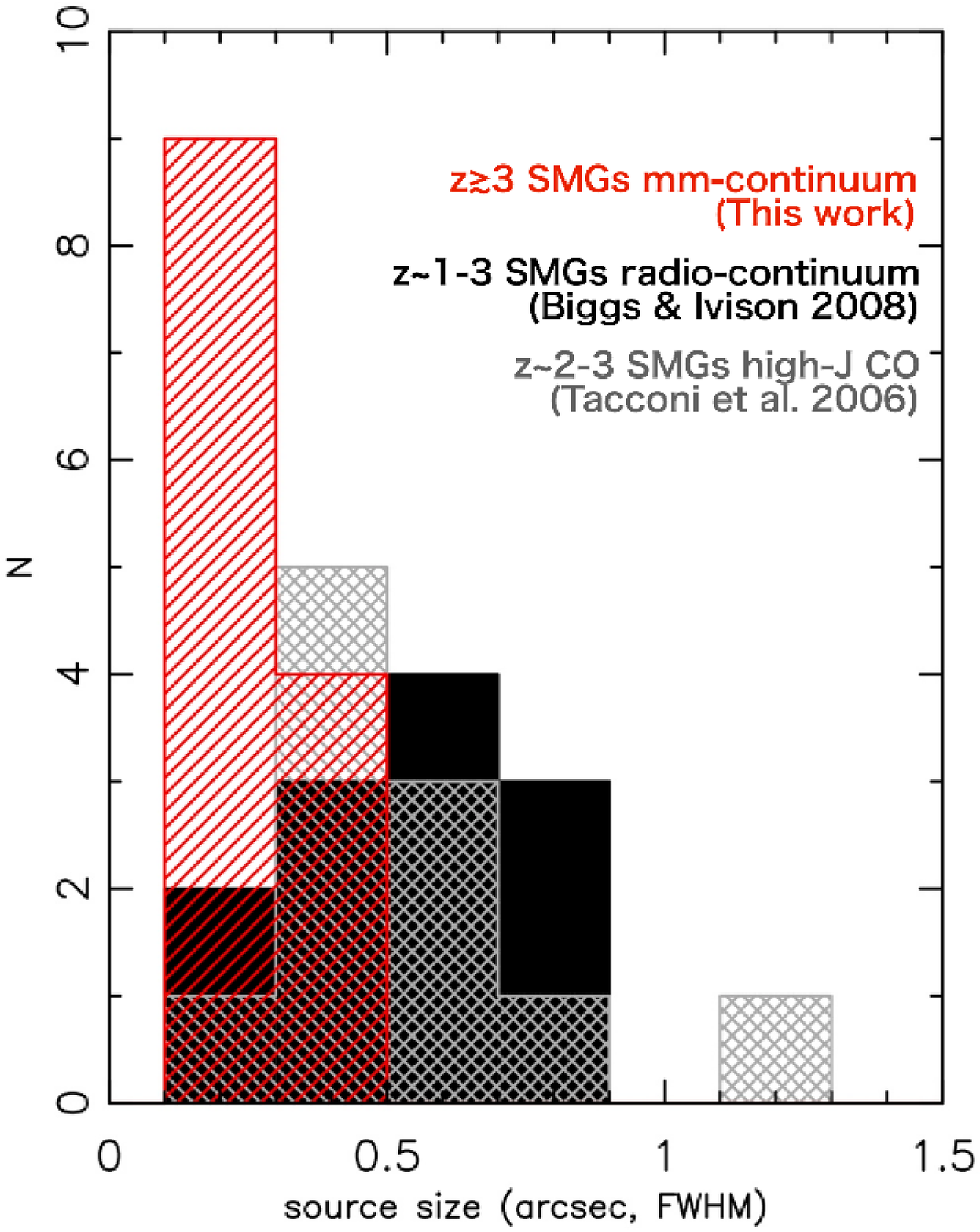}
\vspace{2.0mm}
\caption{Size distribution of $z\gtrsim 3$ SMGs, as measured directly in dust continuum at 1100\,$\mu$m, in comparison with the radio sizes \citep{big08} and CO emission-line sizes \citep{tac06} of $z\sim 1$--3 SMGs. The sizes measured for $z\gtrsim 3$ SMGs are on average about half of those measured in the radio and CO for $z\sim 1$--3 SMGs. In this plot, we have not applied any correction by a possible difference between CO line, millimeter and radio continuum emissions which is discussed in Section\,\ref{sec:size}. 
\label{fig:size}}
\end{center}
\end{figure}

\begin{figure}
\begin{center}
\includegraphics[angle=0,scale=.5]{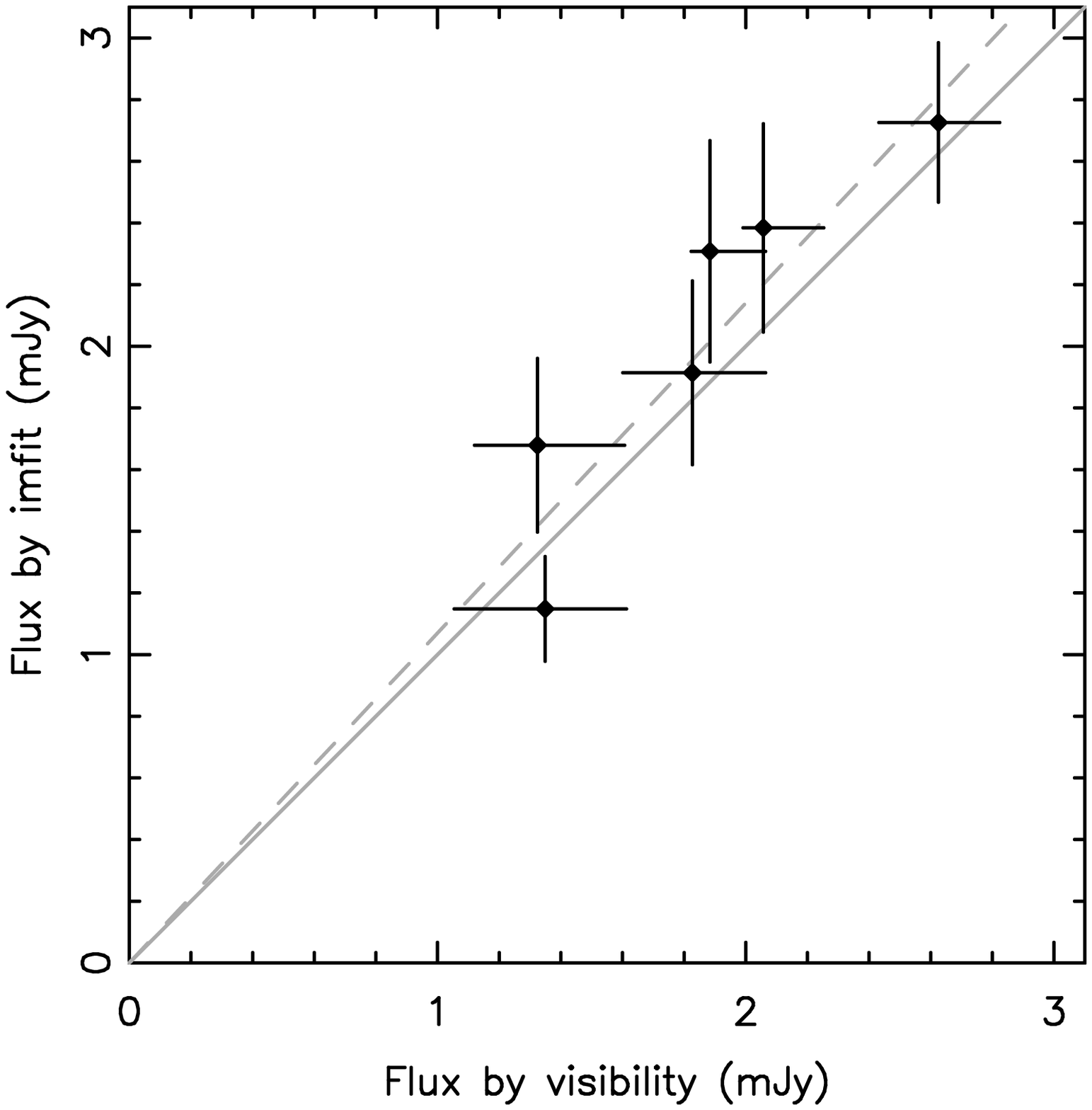}
\vspace{1.0mm}
\caption{
Comparison between flux at $uv$-distance $\geq$ 200\,k$\lambda$ expected from the size by $uv$-amplitude plot and flux measured by imfit on the high-res image. 
Error in flux by visibility comes from the measured size uncertainty shown in Figure~\ref{fig:uvamp}. 
Error in flux by {\sc imfit} is output by imfit. 
The dashed gray line shows flux by {\sc imfit}\,$=$\,1.07$\times$flux by visibility.   
\label{fig:comp}}
\end{center}
\end{figure}

\begin{figure}
\begin{center}
\includegraphics[angle=0,scale=.70]{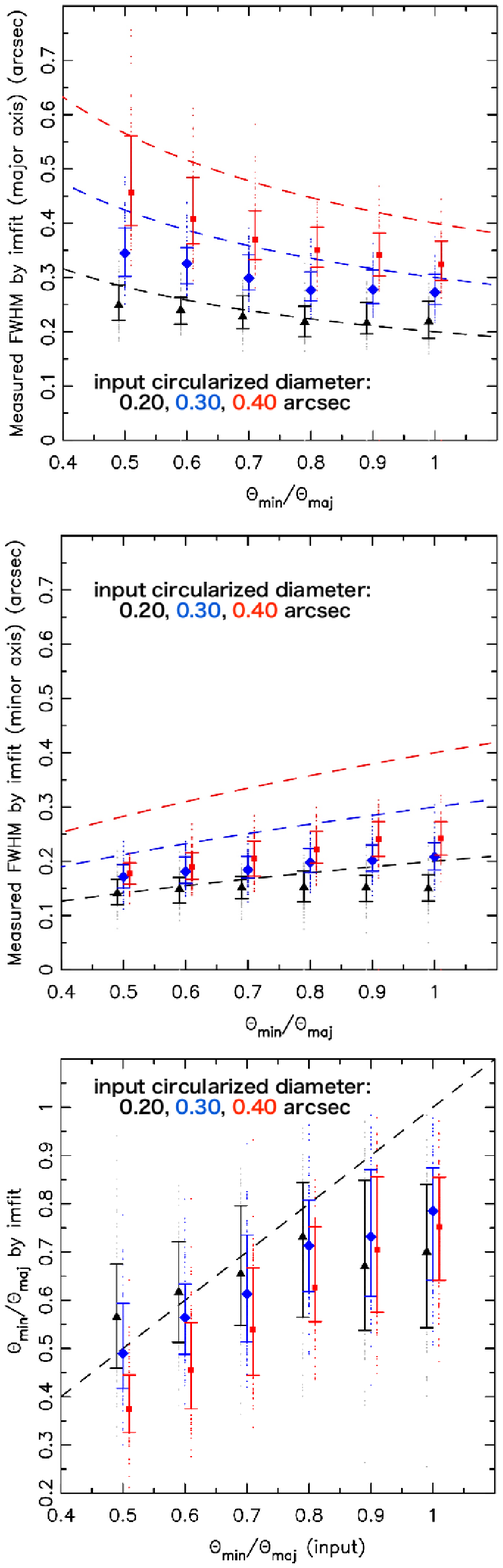}
\vspace{1.0mm}
\caption{
(Top) Relationship between intrinsic minor/major axis ratio and measured major axis size by {\sc imfit} on high-res ALMA image by the Monte Carlo simulation. 
Dashed colored curves are expected sizes of major axis from input visibility model for each size. 
(Middle) Relationship between intrinsic minor/major axis ratio and measured minor axis size by {\sc imfit} on high-res ALMA image by the simulation. 
Dashed colored curves are expected sizes of minor axis from input visibility model for each size. 
(Bottom) Relationship between intrinsic minor/major axis ratio and measured major/minor axis ratio by {\sc imfit} on high-res ALMA image by the simulation. 
We plot adding offset of $-$0.01, 0 and $+$0.01 to $\theta_{\rm min}$/$\theta_{\rm maj}$ for visibility. 
\label{fig:isimu}}
\end{center}
\end{figure}

\section{Are $z\sim3$--6 SMGs more compact than $z\lesssim3$ SMGs?}\label{sec:size}

Our studies have revealed that the typical physical size (median, $R_{\rm c, e}$) of starburst nuclei in $z\sim3$--6 SMGs is 0.67$^{+0.13}_{-0.14}$ kpc (Table~\ref{tbl-1}). 
In the conversion from intrinsic angular source size to physical scale, we assume uniform redshift probability at  $z=3$--6 for sources without photo-$z$ and within $1\sigma$ error of photo-$z$ for the sources with photo-$z$. 
Figure~\ref{fig:size} reveals that our measured sizes are more than a factor $2\times$ smaller than those of SMGs at $z\sim 1$--3 (median radius of 0$''$.5 or 2.5\,kpc as measured via radio continuum -- \citealt{big08} -- or $\sim 0''.5$ or 2\,kpc as measured via high-$J$ CO emission -- \citealt{tac06}).  These radio and CO sizes were measured by Gaussian fitting, as with our measurements, so the comparison is fair.  Errors in calibration cause smearing in interferometric data, but we have no reason to suspect that these larger measurements are due to flawed calibration.  If radio and (sub)millimeter continuum and CO line emission trace star-forming regions, Figure~\ref{fig:size} implies that $z\sim1$--3 and $z\gtrsim 3$ SMGs have different characteristic sizes.  However, local galaxies were also reported to have smaller FIR sizes than CO line(s) \citep[e.g.][]{sak99, sak06, wilc08} and radio continuum \citep[Section 5.1.1 in][]{elb11}.  
Note that the FIR/radio size ratio of 0.86 in local star forming galaxies shown in \citet{elb11} is not sufficient to explain the difference in the millimeter sizes in our sample and the radio sizes in the previous studies.  
In addition, the sizes of radio-detectable SMGs may be affected by radio emission related to radio-loud active galactic nuclei \citep[e.g.][]{ivi10}.   
A Kolmogorov-Smirnov (KS) test gives a probability of 0.3 percent that the differences between radio and millimeter sizes could arise by chance. 
Another KS test with radio size correction due to the empirical FIR/radio size ratio of 0.86 gives 3.5 per cent that the differences between radio and millimeter sizes could arise by chance. 
Given a different scale at $z\sim1$--3 and $z\sim3$--6,  the probability of 3.5 percent is the upper limit, and thus  the difference in the size distributions is significant with $>$ 96.5 percent. 
About CO sizes in Figure~\ref{fig:size}, \citet{big08} presented that the CO sizes in \citep{tac06} and their radio sizes are consistent by a KS-test providing a probability of 84 percent.

Recently sub-millimeter continuum (870 $\mu$m) source size measurements by ALMA of SCUBA2 sources including 23 SMGs with $>$10$\sigma$ detections covering $z\sim1-5$ ($z\sim 3$; median) is also reported  \citep{sim15}. 
Their ALMA data were taken by an array configuration similar to our Schedule Block 1 yielding a median synthesized beam of 0$''$.35$\times$0$''$.25 with the benefit of shorter observing wavelength than ours. 
Their sample consists mainly of SMGs with optical/NIR-detections and photo-$z$ by optical/NIR data contrary to our sample consisting mainly of SMGs faint at optical/NIR wavelength. 
Their sample is typically twice brighter  (5.7$\times$10$^{12}$ $L_{\odot}$; median) than our sample in infrared luminosity. 
They report a median size of 0$''$.30$\pm$0.04 (major axis; FWHM) and 1.2$\pm$0.1\,kpc ($R_{\rm e}$) by Gaussian fitting in images not visibilities.  
The median in SCUBA2 sample ($R_{\rm e}=$ 1.2\,kpc) seems to be $\sim \times1.8$ larger than that in our sample ($R_{\rm c,e}=$ 0.67\,kpc). 
However, we need to take into account the fact that they measured $R_{\rm e}$ of major axis and we measured circularized $R_{\rm c,e}$.
Then we can not compare our sizes with theirs  more in details here, but both of our sample and SCUBA2 sources show smaller FIR continuum sizes of star forming region of SMGs than the previous radio and CO sizes in spite of the different luminosity and redshift distributions in the two samples.   
In order to reveal the possible relation in FIR-continuum size and redshift (and $L_{IR}$), we need higher-angular resolution imaging of SMGs with various properties by ALMA.

\section{$z\gtrsim3$ SMGs as progenitors of the compact quiescent galaxies at $z\sim 2$}

\begin{figure*}
\begin{center}
\includegraphics[angle=0,scale=.50]{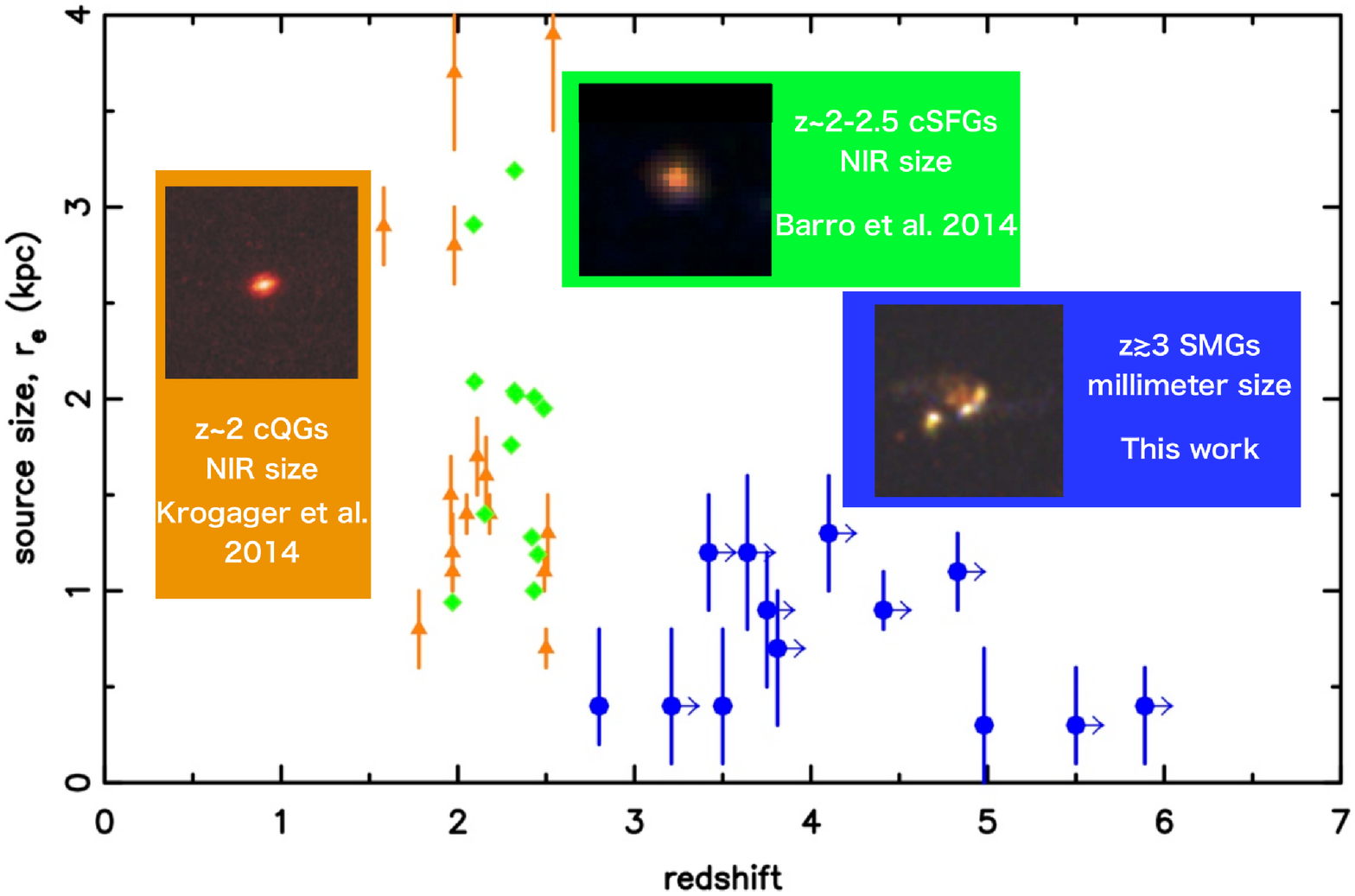}
\vspace{8.0mm}
\caption{Relationship between redshift and sizes for $z\gtrsim 3$ SMGs, $z\sim 2$ cQGs and $z\sim 2$--2.5 cSFGs. 
We plot the 1100-$\mu$m size -- that of the starburst nuclei -- for $z\gtrsim 3$ SMGs (this work).   
We plot the NIR size -- that of the stellar component -- for cQGs \citep{kro13} and cSFGs \citep{bar14} with spectroscopic redshifts. 
Color images of a SMG and a cQGs are taken from \citet{tof14}; that of a cSFG is from \citet{nel14}.  
This plot illustrates that $z\gtrsim 3$ SMGs have a compact starburst region which could generate the compact, high-density stellar components of cQGs or cSFGs. 
Errors in the measured sizes of cSFGs are small ($\sim 0.05$\,kpc) \citep{bar14}. 
\label{fig:sizered}}
\end{center}
\end{figure*}

\citet{tof14} suggested a plausible evolutionary connection between $z\sim 3$--6 SMGs as merger-driven ULIRGs and $z\sim2$ cQGS, based on the following facts: i) the star-formation history of $z\sim2$ cQGs \citep{kro13} matches $z\gtrsim 3$--6 SMGs; ii) the NIR sizes of $z\gtrsim3$ SMGs are compact enough for them to be progenitors of cQGs; iii) simulations suggest that major mergers at $z\gtrsim3$ can generate compact stellar components \citep[e.g.][]{wuy10}.  Our results provide direct evidence that starbursts in $z\gtrsim 3$ SMGs are compact.  We plot the sizes of $z\gtrsim3$ SMGs alongside the NIR sizes of compact star-forming galaxies (cSFGs) at $z\sim 2$--2.5 \citep{bar14} and cQGs \citep{kro13} as a function of redshift in Figure~\ref{fig:sizered}.  The size of the starburst region in $z\sim 3$--6 SMGs is comparable to (or smaller than) the (NIR) size of the stellar component in $z\sim2$ cQGs, supporting the idea that compact dust-obscured starbursts in $z\sim 3$--6 SMGs generate the extremely compact, dense stellar components of cQGs.

Compact star-forming galaxies at $z\sim 2$--2.5 were reported recently as a possible progenitors of cQGs, based on NIR spectroscopy of emission lines and NIR source size measurements, but without direct size measurements of the star-forming region.
cSFGs have similarities with cQGs:  cSFGs have similar NIR structural profiles as cQGs \citep{nel14} and the formation redshift of cSFGs is also similar to that of cQGs \citep{bar14b}. 
These facts could imply that $z\gtrsim3$ SMGs evolve into cQGs via cSFGs (Figure~\ref{fig:sizered}).  
Note that \citet{bar14b} suggests the posibility of disk instability \citep[e.g.][]{dek09, cev10} as another path to make cSFGs and cQGs, based on simulations, in addition to major mergers at $z\gtrsim3$.  Our results make the evolutionary scenario suggested by \citet{tof14} more plausible.    Given that $z\sim2$ cQGs are thought to evolve into local giant ellipticals mostly via dry mergers \citep[e.g.][]{bez09,new12, ose12, kro13}, our results indicate that local giant ellipticals probably experienced an SMG phase at $z\sim3$--6.

\section{Surface star formation rate density of $z\gtrsim3$ SMGs similar to that of local (U)LIRGs}

A discussion based on the surface SFR density ($\Sigma_{\rm SFR}$) is helpful to understand the origin of the compact but huge star-formation activity in $z\gtrsim3$ SMGs. 
We derived $\Sigma_{\rm SFR}$ of our sample using estimated $R_{\rm c, e}$ (Table\,\ref{tbl-1}). 
$\Sigma_{\rm SFR}$ of our sample are in the range of $\sim$30--600\,M$_{\odot}$\,yr$^{-1}$\,kpc$^{-2}$  with a median of 100$^{+42}_{-26}$\,M$_{\odot}$\,yr$^{-1}$\,kpc$^{-2}$. 
We can find that ASXDF sources with a millimeter size of $\sim$\,0$''$.10 (FWHM) show large uncertainty in their $\Sigma_{\rm SFR}$ (Table\,\ref{tbl-1}). 
These large uncertainties in $\Sigma_{\rm SFR}$ come from the large fraction of their size errors to their millimeter sizes which contribute to $\Sigma_{\rm SFR}$ by $\propto R_{\rm c,e}^{-2}$ . 

\begin{figure}
\begin{center}
\includegraphics[angle=0,scale=.45]{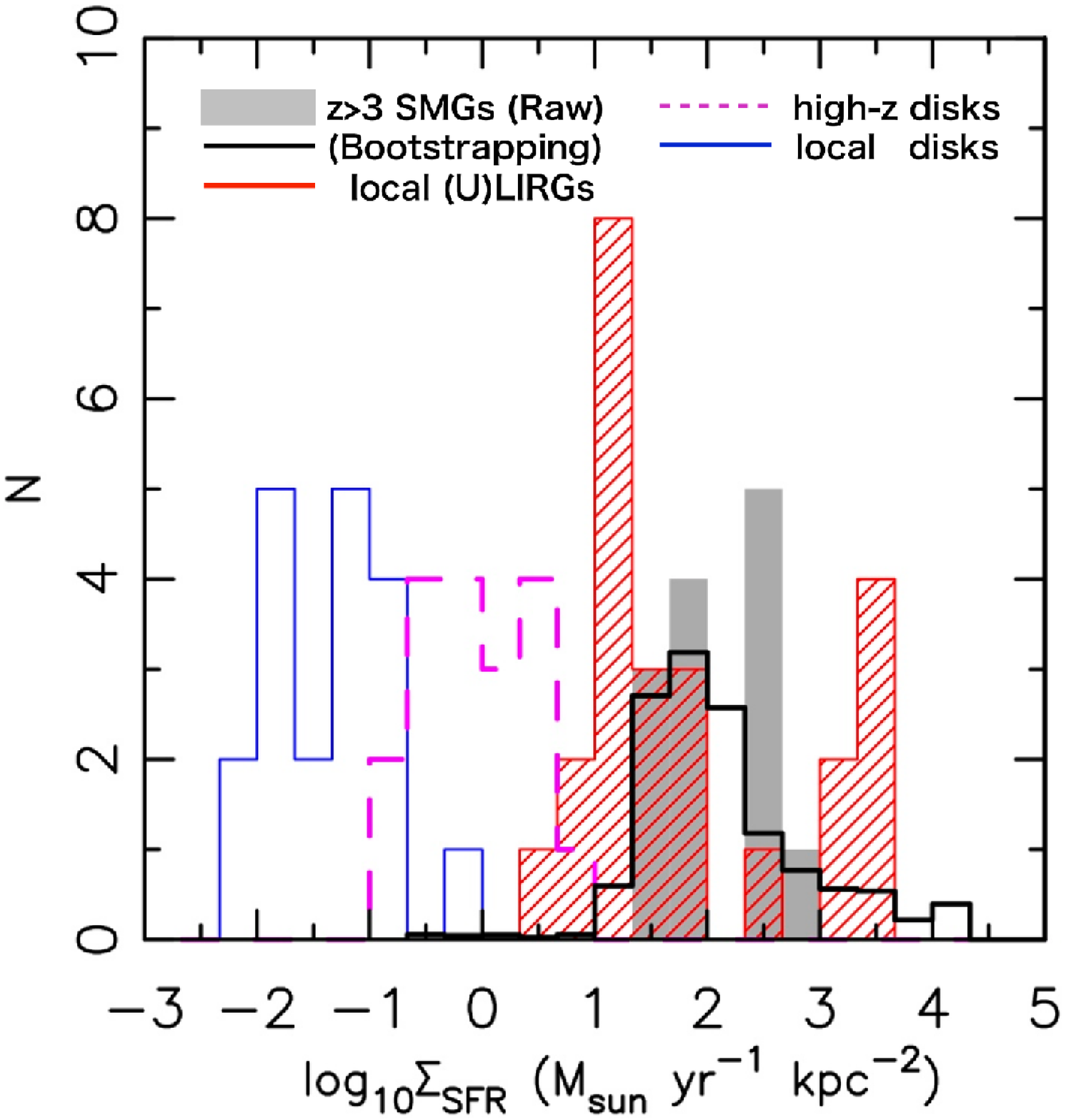}
\vspace{2.0mm}
\caption{Surface star fomation rate density ($\Sigma_{\rm SFR}$) distribution of $z\gtrsim3$ SMGs in comparison with local (U)LIRGs and disk galaxies \citep{ruj11} and high-$z$ extended disks \citep{dad10,tac13}. 
Here, we have not applied any correction of a possible size difference between CO line and (sub)millimeter continuum sizes in high-$z$ disks. 
We derived the expected distribution of $\Sigma_{\rm SFR}$ of $z\gtrsim 3$ SMGs by bootstrapping in order to take into account the large uncertainty of $\Sigma_{\rm SFR}$ of $z\gtrsim 3$ SMGs (Table~\ref{tbl-1}). 
Our $z\gtrsim3$ SMGs show $\Sigma_{\rm SFR}$ distribution similar to local (U)LIRGs. 
\label{fig:ssfrd}}
\end{center}
\end{figure}

First we compare with local galaxies. 
Given that $\Sigma_{\rm SFR}$ for local merger-driven (U)LIRGs is 5--4500\,M$_{\odot}$\,yr$^{-1}$\,kpc$^{-2}$ with a median of 29$^{+24}_{-12}$\,M$_{\odot}$\,yr$^{-1}$\,kpc$^{-2}$ and that  $\Sigma_{\rm SFR}$ for local disks is 0.01--1\,M$_{\odot}$\,yr$^{-1}$\,kpc$^{-2}$ with a median of 0.04$^{+0.02}_{-0.02}$\,M$_{\odot}$\,yr$^{-1}$\,kpc$^{-2}$  \citep{ruj11}\footnote{\citet{ruj11} measured the sizes of local galaxies after convolving high-resolution images to compare with high-redshift sources. We derived $\Sigma_{\rm SFR}$ from the surface infrared luminosity densities of local galaxies in \citet{ruj11} for a Chabrier IMF \citep{cha03}.},  
$z\gtrsim 3$ SMGs are similar to local (U)LIRGs (Figure~\ref{fig:ssfrd}). 
A KS test gives a probability of 3.5 percent that the differences between $\Sigma_{\rm SFR}$ distributions of high-$z$ SMGs and local (U)LIRGs could arise by chance, and thus the two distribution are consistent with  a significant level of 3.5 percent.
The range of the infrared luminosities of local (U)LIRG sample is 10$^{11.1-12.3}$ $L_{\odot}$ which is a little bit fainter than that of our sample. 
It is worth mentioning that a brighter half of local (U)LIRGs with 10$^{11.7-12.3}$ $L_{\odot}$ shows more similar $\Sigma_{\rm SFR}$ distribution to that of high-$z$ SMGs; a KS test gives a probability of 17.1 percent that the differences could arise by chance indicating that the $\Sigma_{\rm SFR}$ distributions of local brighter (U)LIRGs and high-$z$ SMGs are consistent with a significant level of 17.1 percent. 
On the other hand, the remaining fainter local LIRGs show a less similar $\Sigma_{\rm SFR}$ to high-$z$ SMGs; a KS test shows a probability of 0 percent that the differences could arise by chance. 
A KS test also shows that $\Sigma_{\rm SFR}$ distributions of local disks and high-$z$ SMGs do not match with a probability of 0 indicating that star-formation of high-$z$ SMGs is different from  that in local disks. 
Next we compare with high-$z$ extended disk galaxies. 
We took four BzK galaxies at $z\sim$\,1.5 with CO $J$=2--1 sizes from \citet{dad10} and 14 high-redshift disk galaxies at $z\sim1$--2 with CO $J$=3--2 sizes from \citet{tac13} as our sample of high-redshift disk galaxies.  These CO sizes have been derived assuming Gaussian profiles.  
The sample of 18 high-$z$ disk galaxies is among 42 disk galaxies with CO detections in \citet{tac13} including 6 BzKs by \citet{dad10}. The remaining 24 sources tend to be observed by low angular resolutions and are unresolved in CO images. Here we adopt the distribution of CO sizes of the 18 high-$z$ disk galaxies as the representative of the 42 sources because of following facts. \citet{tac13} shows that a ratio of $R_{\rm e}$(optical)/$R_{\rm e}$(CO) is $=1.02\pm0.06$. We checked that optical size distributions of the high-$z$ disk galaxies with  CO sizes and those without CO sizes are consistent with a significant level of 99.7 percent by a KS test.   
The SFRs of these galaxies are $\sim40$--500\,M$_{\odot}$\,yr$^{-1}$ and their sizes ($R_{\rm e}$) range within 2--12\,kpc.  
The distribution of $\Sigma_{\rm SFR}$ for the high-redshift disk galaxies ranges 0.1-7.0\,M$_{\odot}$\,yr$^{-1}$\,kpc$^{-2}$ with a median of 0.5$^{+1.0}_{-0.04}$\,M$_{\odot}$\,yr$^{-1}$\,kpc$^{-2}$. 
A KS test shows that there is a probability of 0 percent that  the differences between $\Sigma_{\rm SFR}$ distributions of high-$z$ SMGs and the high-$z$ extended disks could arise by chance. 

We should take into account a possible difference in CO and (sub)millimeter sizes in high-$z$ SMGs and star-forming galaxies. 
However, we do not know a size correction factor of CO/(sub)millimeter sizes at this moment. 
Then we adopt here CO/(sub)millimeter size ratio of 2.9 and 1.7 derived from CO sizes of SMGs in \citet{tac06} and millimeter sizes in our ASXDF sources, and submillimeter sizes of SMGs in \citet{sim14}, respectively. 
The estimated  $\Sigma_{\rm SFR}$ for the size correction factor of 2.9 is 0.8--59\,M$_{\odot}$\,yr$^{-1}$\,kpc$^{-2}$ with a median of 4\,M$_{\odot}$\,yr$^{-1}$\,kpc$^{-2}$ and one for the size correction factor of 1.7 is 0.3--20\,M$_{\odot}$\,yr$^{-1}$\,kpc$^{-2}$ with a median of 1\,M$_{\odot}$\,yr$^{-1}$\,kpc$^{-2}$.  
Then the $z\gtrsim3$ SMGs seem to have typically $\gtrsim$25$\times$ larger  $\Sigma_{\rm SFR}$ than local and high-$z$ disks. 
KS tests gives probabilities of 0 percent that  $\Sigma_{\rm SFR}$ of high-$z$ extended disks with both of the size correction factor of 2.9 and 1.7 are consistent with that of high-$z$ SMGs.

Figure\,\ref{fig:ssfrd} shows the distributions of $\Sigma_{\rm SFR}$ of our sample in
comparison with local (U)LIRGs, extended disks in high and low redshifts.
We derived the expected distribution of $\Sigma_{\rm SFR}$ of $z\gtrsim 3$ SMGs by bootstrapping in order to take into account the large uncertainty of $\Sigma_{\rm SFR}$ of $z\gtrsim 3$ SMGs (Table\,\ref{tbl-1}). 
This large uncertainty comes from the limit of size measurement in our data set. 
Figure\,\ref{fig:ssfrd}  shows that our sample has $\Sigma_{\rm SFR}$ similar to local (U)LIRGs rather than those of extended disks at high and low redshifts.
Even though the compact star-forming region could consist of more compact clumps spreading in $\leq$0.3--1.3\,kpc as some local ULIRGs show \citep[e.g.][]{sak08,wil14}, almost all of FIR continuum in ASXDF sources come from one compact region as we presented in Section\,3.2. 
Then the fact that  the $\Sigma_{\rm SFR}$ of $z\gtrsim3$ ASXDF sources is similar to that of local (U)LIRGs indicates that $z\gtrsim3$ SMGs also harbor a starburst nuclei as local (U)LIRGs do.

Although local (U)LIRGs are widely thought to be merger-driven, the fact that $z\gtrsim3$ SMGs harbor a compact starburst nuclei as local (U)LIRGs harbor does not imply that the compact starbursts in $z\gtrsim3$ SMGs are triggered by mergers. It is because simulations suggest that disk instability also trigger compact starbursts at high redshift. 
Recently, disk instabilities have been proposed to play important role in triggering starbursts. 
For example, the GALFORM semi-analytic model of galaxy formation including  both types of  starbursts merger-driven and disk instability-driven \citep[][]{lag12,gon14,cow15}, and predicts that disk instabilities are the main channel for triggering starbursts at high-$z$ that would be seen as SMGs \citep{cow15}. Older versions of the same semi-analytic model argued for merger-driven starbursts as the main formation channel of SMGs \citep{bau05}. 
The predicted sizes resulting in starbursts from both disk instabilities and mergers in the GALFORM are in the range $\sim 0.8-2$ kpc ($R_{\rm e}$). 
However, the simplicity of the angular momentum evolution models applied to semi-analytic model prevents us from ruling out the possibility that the high SFR surface densities can only be achieved during galaxy mergers. 
Thus at this stage we cannot distinguish the trigger of starbursts of $z\gtrsim3$ SMGs by their compact size. 
Detailed hydro-dynamic simulations addressing this issue are necessary to shed light into how varied are the progenitors of $z\sim$3--6 SMGs. 
We also need further CO and continuum observations of these galaxies with higher angular resolution and sensitivity by ALMA to reveal the trigger of high-$z$ SMGs. 

Regardless of the triggering mechanism of the $z\gtrsim3$ SMGs, the fact that these are very compact starburst regions supports the evolutionary link between $z\gtrsim3$ SMGs and $z\sim2$ cQGs proposed by \citet{tof14}.

\section{Summary} 

We have exploited new ALMA 1100-$\mu$m continuum data to measure the size of dusty, starburst regions in a sample of 13 SMGs at $z\gtrsim 3$.  The radii of $z\sim 3$--6 SMGs with $L_{\rm IR}\sim2$--$6\times10^{12}$\,L$_{\odot}$ ranges from 0$''$.10 to 0$''$.38 (FWHM) with a median of 0$''$.20$^{+0.03}_{-0.05}$, corresponding to a median circularized effective radius $R_{\rm c, e}$ of 0.67$^{+0.13}_{-0.05}$\,kpc.  Our results demonstrate that the star-forming regions of $z\gtrsim 3$ SMGs are more than 2$\times$ smaller than measured previously using radio continuum and CO data for $z\sim1$--3 SMGs.  Our discovery of compact starbursts in $z\gtrsim 3$ SMGs supports the evolutional scenario proposed by \citet{tof14} wherein $z\gtrsim3$ SMGs evolve into the most massive ellipticals in the local Universe, via dry merging of the cQG population seen at $z\sim 2$, meaning that we can now trace the evolutionary path of the most massive galaxies over a period encompassing $\sim 90$\% of the age of the Universe.

\acknowledgments
This paper makes use of the following ALMA data: ADS/JAO.ALMA\#2012.1.00326.S. ALMA is a partnership of ESO (representing its member states), NSF (USA) and NINS (Japan), together with NRC (Canada) and NSC and ASIAA (Taiwan), in cooperation with the Republic of Chile. The Joint ALMA Observatory is operated by ESO, AUI/NRAO and NAOJ.  Soh Ikarashi is supported by the ALMA Japan Research Grant of NAOJ Chile Observatory, NAOJ-ALMA-0036.  This work was supported by Grant-in-Aid for JSPS Fellows Number 25-10420. Rob Ivison acknowledges support from the European Research Council in the form of Advanced Investigator Programme, COSMICISM, 321302. Yoichi Tamura is supported by JSPS KAKENHI No.~25103503.  This work has been partly supported by Mexican CONACyT research grant CB-2011-01-16729.  Daisuke Iono is supported by the 2015 Inamori Research Grants Program

{\it Facilities:} \facility{ALMA}, \facility{ASTE}.

\clearpage

\begin{table}
\begin{center}
\caption{Summary of ASXDF source size measurements.\label{tbl-1}}
\scriptsize
\begin{tabular}{ c c c c c c c c c c c c}
\tableline \tableline 
 Name                     & R.A.                  & Dec.                 & SNR&$S_{\rm 1100\mu m}$  &  SFR$^{\dagger}$&    $L_{\rm IR}^{\dagger}$           & Photo &  \multicolumn{2}{c}{Size (FWHM)}& $R_{\rm c, e}$$^{\ddagger}$   & $\Sigma_{\rm SFR}$  \\  \cline{9-10}
                               &                          &                        &        &                               &                &                                   &        $z$         &raw        & corrected                      &             & \\
                               &   (J2000)            &        (J2000)     &        &      (mJy)                  &  (M$_{\odot}$\,yr$^{-1}$)  &  ($10^{12}$\,L$_{\odot}$)  &                  & (arcsec)   & (arcsec)  &      (kpc)  & (M$_{\odot}$\,yr$^{-1}$\,kpc$^2$)        \\  \tableline
 \multicolumn{12}{c}{Schedule Block 1 (covering 1200\,k$\lambda$)}  \\  \tableline
ASXDF1100.013.1 & 02:16:45.86 & $-$5:03:47.2 & 18.5 &2.44$\pm$0.13 & 440$^{+40}_{-30}$ & 4.4$^{+0.4}_{-0.3}$ & $>$4.8 & 0.298 & 0.32$^{+0.06}_{-0.06}$ & 1.1$^{+0.2}_{-0.2}$ & 60$^{+36}_{-19}$\\ 
ASXDF1100.027.1 & 02:17:20.95 & $-$5:08:37.2 & 14.8 & 1.91$\pm$0.10 & 400$^{+30}_{-30}$  & 4.0$^{+0.3}_{-0.3}$&    2.80$^{+0.48}_{-0.70}$  & 0.076 & 0.10$^{+0.10}_{-0.08}$ & 0.4$^{+0.4}_{-0.3}$  & $430^{+9500}_{-310}$  \\ 
ASXDF1100.045.1 & 02:18:16.04 & $-$4:54:02.8 & 13.0 &2.02$\pm$0.12 & 360$^{+30}_{-30}$ & 3.6$^{+0.3}_{-0.3}$ & $>$5.5 & $<$0.039$^*$ & 0.10$^{+0.10}_{-0.08}$ & 0.3$^{+0.3}_{-0.2}$ & 610$^{+12000}_{-470}$ \\ 
ASXDF1100.045.2 & 02:18:14.89 & $-$4:54:03.9 & 12.9 &1.86$\pm$0.11 & 330$^{+30}_{-30}$ & 3.3$^{+0.3}_{-0.2}$ & $>$3.4 & 0.307 & 0.31$^{+0.10}_{-0.08}$ & 1.2$^{+0.3}_{-0.3}$ & 39$^{+39}_{-14}$ \\ 
ASXDF1100.049.1 & 02:17:32.86 & $-$4:57:00.8 & 12.3 &1.82$\pm$0.10 & 320$^{+30}_{-20}$ & 3.2$^{+0.3}_{-0.2}$ & $>$3.8 & 0.243 & 0.28$^{+0.10}_{-0.14}$ & 0.9$^{+0.3}_{-0.4}$ & 61$^{+150}_{-29}$ \\ 
ASXDF1100.053.1 & 02:16:48.20 & $-$4:58:59.6 & 27.6 &3.45$\pm$0.10 & 610$^{+50}_{-30}$ & 6.1$^{+0.5}_{-0.3}$ & $>$4.4 & 0.300 & 0.28$^{+0.04}_{-0.04}$ & 0.9$^{+0.2}_{-0.1}$ & 91$^{+40}_{-23}$ \\   
\multicolumn{3}{c}{Stacked faint (ASXDF1100.027.1, 45.1, 45.2 and 49.1)} & 25.0 &1.90$\pm$0.05 & 340$^{+30}_{-20}$ & 3.4$^{+0.3}_{-0.2}$ & --- & 0.170 & 0.18$^{+0.06}_{-0.10}$ & 0.6$^{+0.2}_{-0.3}$  & 160$^{+580}_{-72}$\\ 
\multicolumn{3}{c}{Stacked all (faint $+$ ASXDF1100.013.1 and 053.1)} & 31.0 &2.37$\pm$0.03 & 420$^{+40}_{-20}$ & 4.2$^{+0.3}_{-0.2}$ & --- & 0.240 & 0.24$^{+0.04}_{-0.06}$ & 0.8$^{+0.2}_{-0.2}$ & 110$^{+68}_{-32}$ \\ 
\tableline
 \multicolumn{12}{c}{Schedule Block 2, 3 (covering 400\,k$\lambda$)}  \\  \tableline
ASXDF1100.073.1 & 02:18:10.04 & $-$5:11:31.7 & 10.4 &1.23$\pm$0.07 & 220$^{+20}_{-20}$ & 2.2$^{+0.2}_{-0.2}$ & $>$3.6 & 0.308 & 0.34$^{+0.10}_{-0.12}$ & 1.2$^{+0.4}_{-0.4}$ & 26$^{+38}_{-11}$  \\ 
ASXDF1100.083.1 & 02:17:12.42 & $-$5:03:59.4 & 13.7 &2.03$\pm$0.08 & 360$^{+30}_{-20}$ & 3.6$^{+0.3}_{-0.2}$ & $>$4.1 & 0.355 & 0.38$^{+0.08}_{-0.06}$ & 1.3$^{+0.3}_{-0.3}$  & 34$^{+21}_{-10}$  \\ 
ASXDF1100.090.1 & 02:17:23.04 & $-$4:57:29.9 & 11.3 &1.62$\pm$0.11 & 290$^{+30}_{-20}$ & 2.9$^{+0.3}_{-0.2}$ & $>$3.2 & $<$0.039$^*$ & 0.10$^{+0.12}_{-0.08}$ & 0.4$^{+0.4}_{-0.3}$ & 360$^{+9100}_{-275}$   \\ 
ASXDF1100.110.1 & 02:17:43.59 & $-$5:04:10.3 & 12.7 & 1.46$\pm$0.08 & 275$^{+40}_{-30}$ & 2.8$^{+0.4}_{-0.3}$ &    4.98$^{+0.72}_{-3.14}$ & $<$0.039$^*$ & 0.10$^{+0.10}_{-0.08}$ & 0.3$^{+0.4}_{-0.3}$ & 420$^{+7800}_{-330}$  \\ 
ASXDF1100.127.1 & 02:17:33.36 & $-$4:48:43.8 & 13.3 &1.81$\pm$0.10 & 320$^{+30}_{-20}$ & 3.2$^{+0.3}_{-0.2}$ & $>$3.8 & 0.192 & 0.22$^{+0.16}_{-0.14}$ & 0.7$^{+0.3}_{-0.4}$ & 99$^{+570}_{-50}$   \\ 
ASXDF1100.230.1 & 02:17:59.39 & $-$4:45:53.1 & 11.3 & 1.86$\pm$0.13 & 350$^{+20}_{-30}$ & 3.5$^{+0.2}_{-0.2}$ &    3.50$^{+0.40}_{-0.18}$  & $<$0.039$^*$ & 0.10$^{+0.12}_{-0.08}$ & 0.4$^{+0.4}_{-0.3}$ & 380$^{+9500}_{-300}$   \\ 
ASXDF1100.231.1 & 02:17:59.65 & $-$4:46:49.7 & 26.8 &2.27$\pm$0.06 & 400$^{+30}_{-20}$ & 4.0$^{+0.3}_{-0.2}$ & $>$5.9 & 0.107 & 0.12$^{+0.08}_{-0.08}$ & 0.4$^{+0.2}_{-0.3}$ & 450$^{+3600}_{-270}$   \\ 
\tableline 
\end{tabular} 
\tablenotetext{}{\selectfont All values in this table are measured on ALMA data not on AzTEC data.} 
\tablenotetext{$\dagger$}{$L_{\rm IR}$ and SFR assume average SED of ALMA-identified SMGs \citep{swi14}, with a Chabrier IMF \citep{cha03}. We also assume uniform redshift probability at $z=3$--6 for sources without photo-$z$, and in 1$\sigma$ error for sources with photo-$z$.}
\tablenotetext{$\ddagger$}{$R_{\rm c, e}$ is derived from the half width at half maximum (HWHM) assuming a symmetric Gaussian profile. HWHM corresponds to $R_{\rm c, e}$ in a symmetric Gaussian profile. We also assume a same redshift probability as we do for $L_{\rm IR}$ and SFR. } 
\tablenotetext{*}{Our fitting stops at 0$''$.039 meaning that these sources are unresolved.  Our data does not have enough sensitivity to measure such small sizes and the sizes are determined by simulations in these cases.} 
 \end{center}
\end{table}

\end{document}